\documentclass[namedreferences,hyperref,optionalrh,solaromanenum]{spr-sola}

\usepackage{graphicx}
\usepackage{color}
\usepackage{breakurl}

\chardef\us=`\_

\usepackage{amsmath}
\usepackage{amssymb}
\usepackage{physics}
\usepackage{gensymb}

\usepackage{ltablex}
\usepackage{tabularx}
\usepackage{enumitem}
\newcommand{\sadcatlink}{\url{https://trestansimon.github.io/sadcat-web/}}
\newcommand{\figscale}{0.55}

\begin{document}

\begin{frontmatter}

\title{SADCat: A Catalog of Supra-Arcade Downflow Events in Solar Flares}

\author[addressref={aps,lasp},email={trestan.simon@colorado.edu},corref]{\inits{T.F.}\fnm{Trestan~F.}~\snm{Simon}\orcid{0009-0000-3029-8619}}
\author[addressref={lasp},email={ryan.french@lasp.colorado.edu}]{\inits{R.J.}\fnm{Ryan~J.}~\snm{French}\orcid{0000-0001-9726-0738}}

\runningauthor{T.F. Simon, R.J. French}
\runningtitle{SADCat: A Catalog of Supra-Arcade Downflow Events in Solar Flares}

\address[id=aps]{Department of Astrophysical and Planetary Sciences, University of Colorado Boulder, Boulder, CO 80303, USA}
\address[id=lasp]{Laboratory for Atmospheric and Space Physics, University of Colorado Boulder, Boulder, CO 80303, USA}

\begin{abstract}
Supra-arcade downflows (SADs) are sunward-traveling features routinely observed in hot fan structures above magnetic loop arcades during eruptive solar flares.
We manually compiled a catalog (SADCat) of 178 SAD-productive flares imaged by the Atmospheric Imaging Assembly (AIA) aboard the {Solar Dynamics Observatory} ({SDO}) during solar cycle 24 as a resource for the wider solar flare community.
We conducted a preliminary analysis of the SADCat, comparing the flare X-ray and CME properties between eruptive solar flares with and without SADs.
We found that peak GOES X-ray flux, flare duration, CME speed, and CME mass have a strong influence on whether a flare produces visible SADs, whereas flare impulsivity and CME acceleration have little effect.
\end{abstract}

\keywords{Solar Flares, Coronal Mass Ejections, Database}

\end{frontmatter}

\section{Introduction}

Solar flares and coronal mass ejections (CMEs) are thought to be driven by rapid, large-scale magnetic reconnection in the solar corona, which rearranges the coronal magnetic topology and converts stored magnetic energy into thermal and kinetic energy. In the standard model of eruptive flares \citep[also known as the CSHKP model after][]{carmichael_1964, sturrock_1966, hirayama_1974, kopp_1976}, this reconnection occurs within a vertically oriented current sheet that forms between oppositely directed magnetic fields brought together below a rising flux rope structure. As the magnetic field reconnects, an arcade-like system of closed magnetic loops develops in the corona below the current sheet and emits strongly in soft X-ray (SXR) and extreme ultraviolet (EUV). Although these current sheets cannot be observed directly, wispy fan-like structures extending above and lengthwise-across the arcade are routinely observed in SXR and EUV imagery \citep{svestka_1998, mckenzie_1999, mckenzie_2000} and are believed to encase the much-narrower current sheet. These supra-arcade fans are filled with turbulent $\sim\hspace{-3pt}10$~MK plasma and can persist for several hours, reaching heights of up to 150~Mm above the arcade looptops \citep{innes_2014, hanneman_2014}. Supra-arcade fans are also frequently interpreted as the face-on manifestation of plasma sheet structures observed in off-limb EUV imagery \citep[e.g.][]{Warren2018, French2020} with edge-on thicknesses of $\sim\hspace{-3pt}3$--30~Mm \citep{longcope_2018}.

Sunward-propagating features referred to as supra-arcade downflows (SADs) are commonly identified in images of supra-arcade fans during eruptive solar flares close to the limb. SADs often appear as dark, blob-like voids with elongated trailing regions and were first identified in SXR imagery from the {Yohkoh} Soft X-ray Telescope \citep[SXT;][]{Tsuneta1991} by \citet{mckenzie_1999}. Since then, SADs have been regularly detected by many other instruments, including in SXR imagery from the {Hinode} X-ray Telescope \citep[XRT;][]{golub_2007} by, for example, \citet{savage_2010}; in EUV spectra from the {Solar and Heliospheric Observatory} (SOHO) Solar Ultraviolet Measurements of Emitted Radiation \citep[SUMER;][]{Wilhelm1995} by \citet{innes_2003} and the Hinode EUV Imaging Spectrometer \citep[EIS;][]{culhane_2007} by \citet{French2025}; in EUV imagery from the Transition Region and Coronal Explorer \citep[TRACE;][]{Handy1999} by \citet{innes_2003} and the {Solar Dynamics Observatory} \citep[SDO;][]{pesnell_2012} Atmospheric Imaging Assembly \citep[AIA;][]{lemen_aia_2012} by \citet{warren_2011}; and in white-light imagery from the {SOHO} Large Angle and Spectrometric Coronagraph Experiment \cite[LASCO;][]{brueckner_1995} by \citet{savage_2011}. The large volume of high-cadence full-disk EUV imagery from AIA has provided the majority of SAD observations and has been instrumental in advancing our understanding of the phenomenon.

Previous studies have found SADs to be regions of low-density \citep{innes_2003, savage_2012}, turbulent plasma \citep{Shen2022, Xie2025, French2025}, with temperatures close to or cooler than the surrounding flare fan \citep{hanneman_2014, Xie2023, French2025}. They have been observed to travel at speeds of 45--500~km~s$^{-1}$ \citep{mckenzie_1999, McKenzie2009, warren_2011, xie_2022}, lower than the local Alfv\'{e}n speed, and most frequently appear during the decay phase of a solar flare \citep{warren_2011}. Many questions about the origin and nature of these downflows remain open, and there are several competing interpretations. The general consensus is that SADs are either directly or indirectly related to magnetic outflows beneath a solar flare current sheet. But, in these scenarios, SADs have been postulated to be either the contracting low-density magnetic loops themselves \citep[e.g.,][]{savage_2011,warren_2011} or regions of low-density plasma in the wake of narrow contracting loops \citep[e.g.,][]{savage_2012, Scott2016}.

Most AIA studies of SADs examine a small number of flares ($\lesssim10$), often investigating properties of individual SAD structures within those events. Prior to the availability of AIA data, there have been several published efforts to catalog longer lists of solar flares producing SADs \citep{mckenzie_2000, khan_2007, savage_2011}. However, to the best of our knowledge, a comprehensive catalog of SAD-productive flares in AIA data has not been presented in the literature. In this work, we describe the methods used to compile a SAD catalog (SADCat) of 178 SAD-associated eruptive flares observed by {SDO}/AIA throughout solar cycle 24. This is a list of eruptive flares that produced SADs, and not a list of individual SAD-features within single events. We present this catalog as a resource for use by the solar flare community in future SAD studies. We further present an initial analysis of the SADCat, comparing the incidence of SADs in flares with SXR and CME properties observed by {Geostationary Operational Environmental Satellite} ({GOES}) and SOHO/LASCO, respectively. Such comparisons provide a rich discovery space to probe the mechanisms driving SAD formation and visibility. We do not aim to provide all of the answers in this study, but provide the SADCat as a resource meant to facilitate future SAD studies, whilst presenting an initial analysis of the relationship between SADs and flare/CME behavior -- some of which are unintuitive. Future theoretical and computational explanations of SADs must explain these relationships presented in this work.

In Section~\ref{sec:procedure} we detail the procedure used to identify SAD events and compile the SADCat; in Section~\ref{sec:results} we discuss the data provided in the SADCat; in Section~\ref{sec:discussion} we present the results of our preliminary analysis; and in Section~\ref{sec:summary} we summarize the results.

\section{Procedure} \label{sec:procedure}
We have assembled a catalog of SAD events observed by {SDO}/AIA during solar cycle~24. Our procedure for constructing the SADCat involved programmatically generating a list of candidate eruptive flares and manually identifying the flares during which SADs were present. Constructing the list of candidate events involved filtering an initial list of SXR flares by their GOES classes and source-region longitudes, and determining the flares' association with a CME. SDO/AIA imagery for each candidate event was then manually inspected to identify events during which SADs were (or were not) visible. This selection procedure is outlined in this section. We have limited our search to events from 2010 May~13 to 2019 December~31, the first date for which Level-1 AIA data is available and the final date of solar cycle~24, respectively. This time range covers approximately 87\% of solar cycle~24 and all of the solar maximum period where flares were most common. 

\subsection{Initial Candidate Flares} \label{ssec:candidate}
The initial list of candidate solar flares was derived from the Hinode Flare Catalog \citep{watanabe_hinode_2012}, a curated and accessible list of {GOES} SXR flares spanning our time interval of interest.
For each flare, the catalog includes the start, peak, and end times; GOES classes; and heliographic Stonyhurst (HS) coordinates obtained from the SolarSoft Latest Events Archive\footnote{\url{https://www.lmsal.com/solarsoft/latest_events_archive.html}} maintained by the Lockheed Martin Solar and Astrophysics Laboratory (LMSAL).
Using these data, we filtered candidate flares down to those with a peak flux in the {GOES} 1--8-\AA\ channel greater than or equal to $10^{-6}\ \text{W m}^{-2}$ (corresponding to a {GOES} class of C1.0) after removal of the SWPC recalibration factor\footnote{\url{https://www.ngdc.noaa.gov/stp/satellite/goes/doc/GOES_XRS_readme.pdf}} and those originating from a region with unsigned HS longitude greater than or equal to 65\degree.
Out of the 14,781 flares listed in the Hinode catalog during our target time range, 3552 satisfied both of these conditions.

\begin{figure}
\centerline{\hspace*{0.015\textwidth}\includegraphics[scale=\figscale]{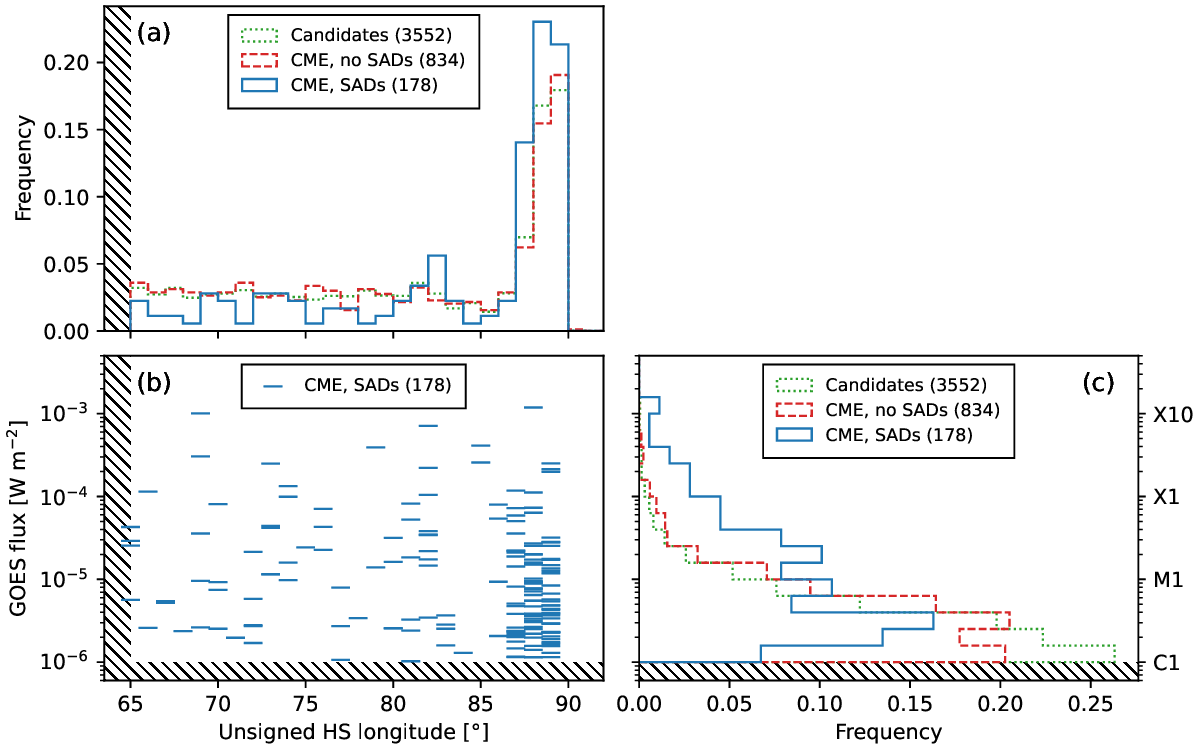}}
\caption{Frequency distributions of (a) the source-region unsigned HS longitudes and (c) the {GOES} fluxes for candidate flares satisfying the initial time, longitude, and flux constraints (green dotted line) and CME-associated flares not associated with SADs (dashed red line) and associated with SADs (solid blue line). (b) Scatter plot of the source-region unsigned HS longitudes and {GOES} fluxes of only CME-associated flares with SADs. Longitude and flux ranges that were not considered in this analysis are indicated with hatched lines.}
\label{fig:flux-lon}
\end{figure}

The frequency distribution of the unsigned HS longitudes and flare classes for the 3552 qualifying flares are shown in Figure~\ref{fig:flux-lon}(a) and~(c), respectively, as a dotted green histogram. These longitudes, labeled the ``Derived Position'' (paired with latitude) in the original LMSAL archive, are the approximate on-disk coordinates of the flare source. The coordinates accurately represent the source location of flares with an unsigned longitude less than about $85\degree$ but are not accurate for flares originating on or behind the limb. This bias assigns a disproportionate number of flares with a longitude of 85--90\degree, which is apparent from Figure~\ref{fig:flux-lon}(a). Despite this limitation, we find that these coordinates are still suitable for filtering out flares near center-disk where SADs are rarely visible.

The constraints applied to the initial list of candidate flares were imposed in order to eliminate flares with properties unlikely to produce clearly observable SADs. The constraint on flare class was chosen with the expectation that flares with a higher peak flux are more likely to produce prominent supra-arcade fans and downflow signatures therein. As such, this catalog neglects flares with little to no observed SXR emission, which can occur when flares are significantly occulted by the limb \citep[e.g.,][]{savage_2010}. Furthermore, the constraint on source-region longitude was chosen with the expectation that fans and SADs are more frequently observed at or near the limb. Although on-disk SADs have been investigated in the literature \citep[e.g.,][]{samanta_plasma_2021, awasthi_effects_2022} and were observed when identifying events for this catalog, past studies have noted that bright on-disk background features may inhibit accurate identification of individual SADs \citep{chen_thermodynamics_2017}. Additionally, because supra-arcade fans often extend radially from the flare arcade, flares originating from source regions with low unsigned longitude produce line-of-sight issues when attempting to identify SADs.

\subsection{Flare--CME Correlation}
Another constraint placed on candidate flares was the requirement that they be eruptive, as SADs have exclusively been associated with eruptive flares. We considered a flare ``eruptive'' if it correlated with a CME documented in the {SOHO}/LASCO CME Catalog from the CDAW Data Center \citep{gopalswamy_2009, gopalswamy_soho_2025}. This CME catalog includes CMEs observed by the LASCO C2 coronagraph for about 97\% of our 2010 May~13 to 2019 December~31 time window. To formulate the conditions under which a flare and CME would be considered correlated, we consulted the past investigations into this correlation using the CDAW catalog in \citet{aarnio_11} and \citet{compagnino_17}. We required that the angular separation between the flare's source region and the CME's central position angle with respect to center disk be less than or equal to either half of the CME's angular width or 45\degree, and, keeping in mind the relative timing of SAD events identified previously in the literature, we required that the CME be detected between the flare's start time and 80~minutes after its peak time. When multiple CMEs satisfied these conditions for any one flare, we chose to use the CME with the smallest angular separation from the flare's source region. However, multiple flares correlating with a single CME was permitted. Applying these conditions to the list of 3552 candidate flares, 1015 flares correlated with one of 921 unique CMEs from the CDAW catalog.

\subsection{Identifying SADs}\label{ssec:id_SADs}

\begin{figure}
\centerline{\hspace*{0.015\textwidth}\includegraphics[scale=\figscale]{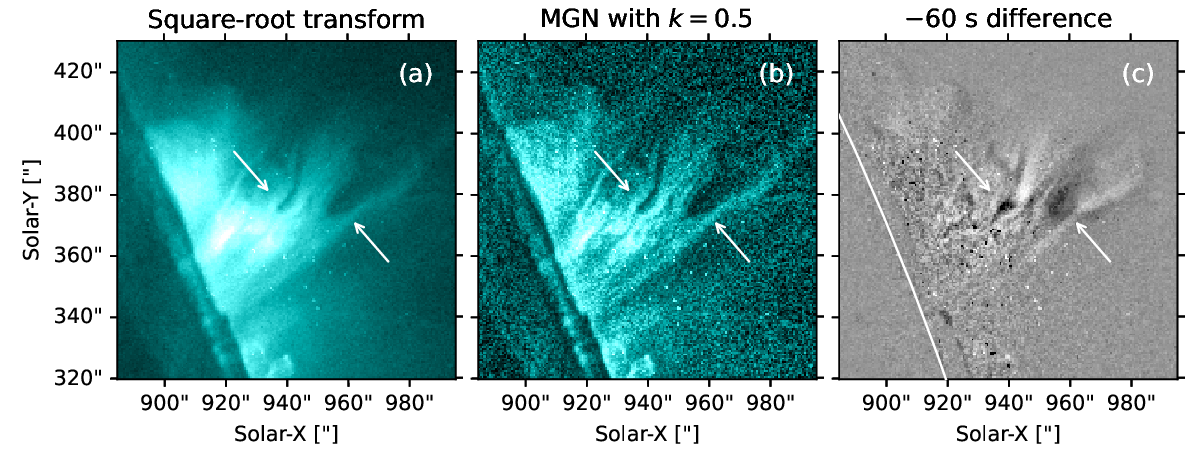}}
\caption{AIA 131-\AA\ imagery from 2011 December~3 at 11:50:02 UTC of SADs processed using the three different techniques mentioned in Section~\ref{ssec:id_SADs}: (a) square-root transformation, (b) multi-scale Gaussian normalization with $k=0.5$, and (c) the difference with the frame taken 60~seconds before.}
\label{fig:aia_single_example}
\end{figure}

We manually identified SADs associated with the candidate events by fetching relevant {SDO}/AIA data from the Joint Science Operations Center using SunPy's drms module \citep{glogowski_2019}. This research used version 6.0.1 \citep{mumford_sunpy_2024} of the SunPy open source software package \citep{sunpy_community2020}. For each event, we acquired a series of $400''\times 400''$ cutouts of Level-1 AIA 131-\r{A} data spanning from the event's start time to 120~minutes after its end time with a 12-second cadence. The initial cutouts were centered on the flares' source-region coordinates and later repositioned or expanded if the supra-arcade fan was not in-frame. Three sets of images were generated from these cutouts by applying square-root transformations, multi-scale Gaussian normalization \citep{morgan_2014}, and time-difference techniques. Figure~\ref{fig:aia_single_example} shows a flare (with associated SADs) in AIA 131-\r{A} imagery after the application of each of the three image processing techniques. Each set of frames was used to create individual movies for each flare, which were then inspected manually to determine whether SADs were present. For events with flaring activity appearing to originate from a source region out-of-frame, the event's source-region coordinates were corrected, and the flare--CME correlation was reevaluated. Flares not fully visible in AIA 131-\r{A} imagery due to an eclipse or calibration maneuver were removed from the list of candidate events; three flares were found to be completely obscured, leaving a total of 1012 candidate events.

Manual inspection of the generated movies for all 1012 candidate events was carried out over a period of several weeks.
SADs were identified in these movies as tadpole- or teardrop-like voids traveling through a supra-arcade fan toward post-flare arcade looptops. In order to maintain consistency with how SADs were identified, a single author completed the identification process.
Supra-arcade downflowing loops \citep[SADLs;][]{savage_2011}, a related phenomenon involving the motion of shrinking loops in the supra-arcade fan, were considered out of the scope of this catalog.

SADs were visible in supra-arcade fans at all viewing angles except in fans viewed fully edge-on. The flare and CME parameters we examined are all independent of the viewing angle of the associated supra-arcade fan, so we do not expect the viewing angle to influence the statistical relationships presented in our preliminary analysis.

\begin{figure}
    \centerline{\hspace*{0.015\textwidth}\includegraphics[scale=0.6]{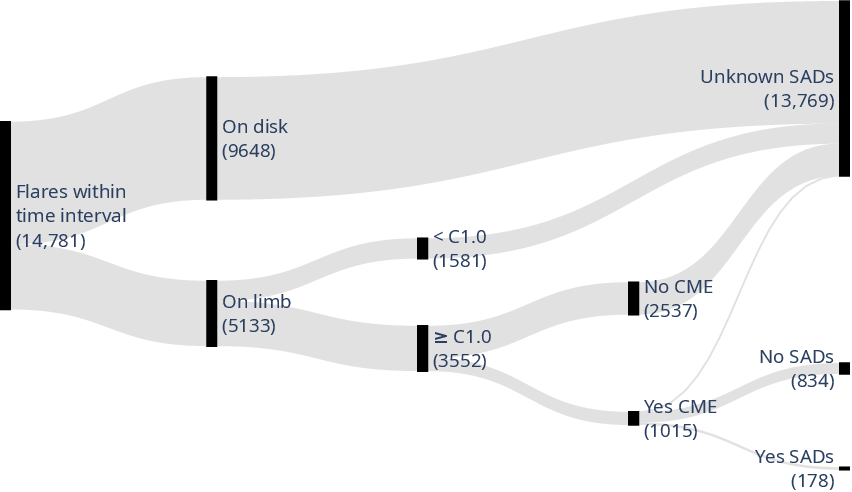}}
    \caption{Sankey diagram of the procedure used to construct the catalog of 178 SAD events starting from the 14,781 flares listed in the Hinode Flare Catalog for our time interval of interest.}
    \label{fig:sankey}
\end{figure}

This identification process ultimately produced a list of flares with visible SADs (SADCat; 178 events, 18\%) and a list of flares without visible SADs (834 events, 82\%). Figure~\ref{fig:sankey} shows a Sankey diagram of this process, reducing the full flare list down to the final SADCat. Figure~\ref{fig:flux-lon}(a) and~(c) show the frequency distribution of the unsigned HS longitudes and flare class, respectively, of SAD flares (solid blue line) and non-SAD flares (dashed red line) in addition to the total 3552 candidate flares (dotted green line). Figure~\ref{fig:flux-lon}(b) shows the longitude versus flare class distribution of SAD flares exclusively. Of the 178 SAD events, 62\% (110) have unsigned HS longitude $\geq85\degree$ compared to 46\% (1633) of the 3552 candidate flares, indicating that SADs are somewhat more likely to be visible close to the limb. This is consistent with our decision to only consider candidate flares above 65\degree\ unsigned longitude, although it should be noted that at least one flare with visible SADs occurred below this threshold during solar cycle 24 \citep{samanta_plasma_2021}. Likewise, of the 178 SAD events, 40\% (71) are of class $\leq\rm C5$ compared to 76\% (2705) of the 3552 candidate flares, indicating that SADs are less likely to be visible near our flux threshold. This supports are decision to include only flares above C1 class, although some exceptionally weak or occulted flares may lie below this threshold.

\section{Catalog of SAD Events} \label{sec:results}

\begin{figure}
\centerline{\hspace*{0.015\textwidth}\includegraphics[scale=\figscale]{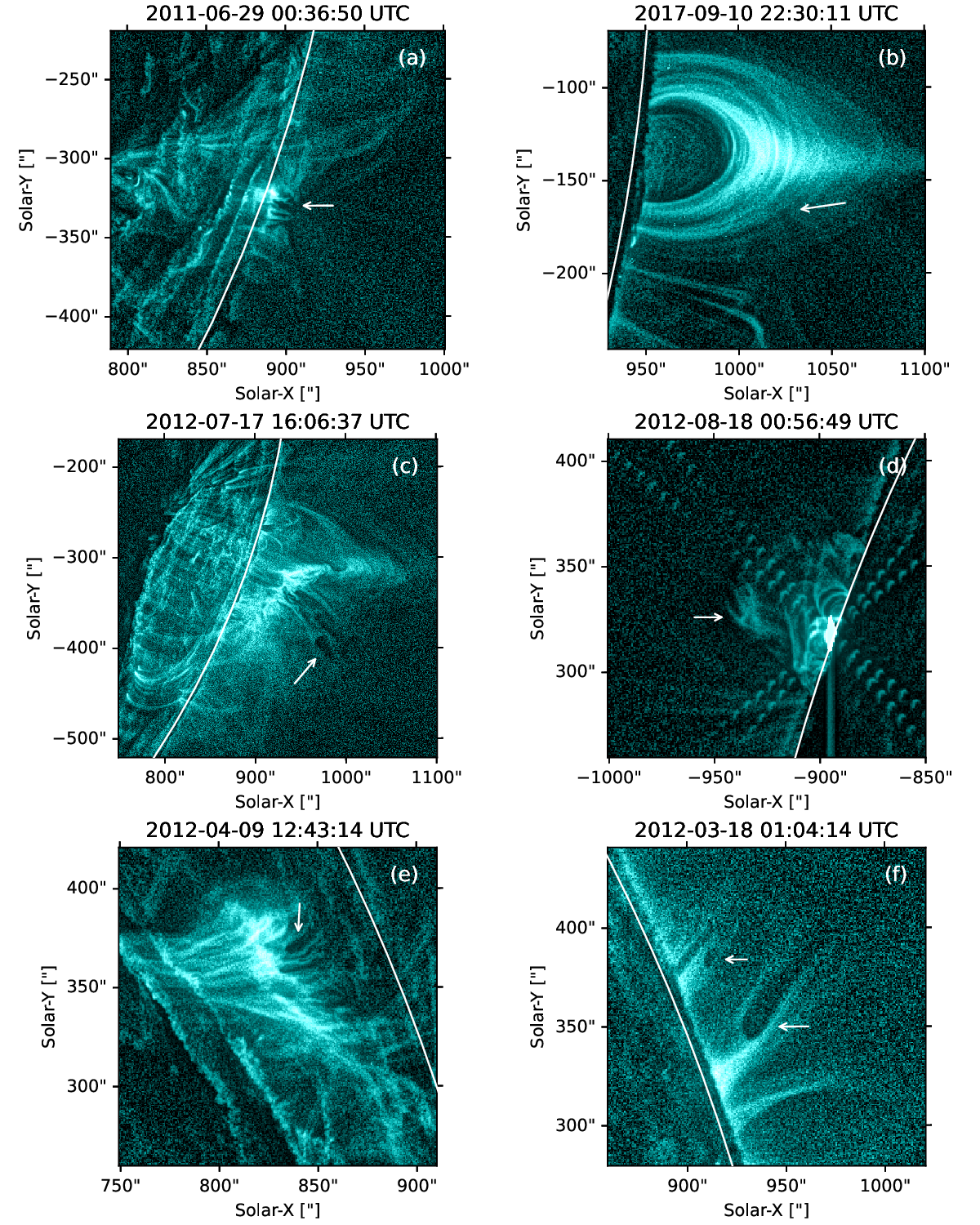}}
\caption{AIA 131-\AA\ imagery of the SAD events with (a) the lowest peak {GOES} flux (C1.04); (b) the largest peak {GOES} flux (X11.7), CME speed (3,163~km~s$^{-1}$), and CME mass ($4.8\times 10^{16}$~g); (c) the longest duration (421~minutes); (d) the greatest impulsivity (0.88); (e) a source region with low unsigned HS longitude; and (f) a source region with high unsigned HS longitude.
}
\label{fig:aia}
\end{figure}

The SADCat is provided in Table~\ref{tbl:sads1} and an online version is available at \sadcatlink. For each SAD event in the catalog, we provide GOES SXR flare, source-region, and CME characteristics. From left to right in Table~\ref{tbl:sads1}, we give the flare's time of peak flux (Peak time) in Coordinated Universal Time, duration (Dur.) in minutes, impulsivity (Imp.), and peak flux class (Class); the source region's active region number (No.) assigned by the National Oceanic and Atmospheric Administration (NOAA) when applicable and its HS latitude (Lat.) and longitude (Lon.) in degrees prefixed by the hemisphere; and the CME's speed (Speed) in kilometers per second and mass (Mass) in grams. Figure~\ref{fig:aia} shows a selection of SAD flares, at the extremes of some of these characteristic values.

The flare peak times/classes and source-region longitudes/latitudes provided in the catalog, in addition to the flare start and end times used to derive the provided durations and impulsivities, were obtained directly from the Hinode Flare Catalogue. The logic defining GOES flare start and end times is provided in the User's Guide for GOES-R XRS L2 Products.\footnote{\url{https://data.ngdc.noaa.gov/platforms/solar-space-observing-satellites/goes/goes16/l2/docs/GOES-R_XRS_L2_Data_Users_Guide.pdf}} Flare durations were found by taking the difference between the start and end times, and impulsivities were calculated from the ratio between the start-to-peak times and flare durations. The {GOES} classes were modified to remove the SWPC factor early in the procedure. Additionally, the source-region coordinates are subject to the limitations discussed in Section~\ref{ssec:candidate}.

The active region numbers assigned by NOAA were obtained programmatically by querying the Heliosphere Event Knowledgebase \citep[HEK;][]{hurlburt_2012} API for active regions in the vicinity of the source region at the flare start time. For SAD-associated source regions that did not automatically receive an active region number, a manual search was conducted to assign a number. The number was set to ``N/A'' if the flare appeared to originate from an unnumbered region.

The CME parameters were obtained directly from the CDAW catalog. The CME speeds (listed under ``linear speed'' in the CDAW catalog) and accelerations were calculated by fitting the height-time measurements made in the LASCO field of view to first- and second-order polynomials, respectively. As such, these values correspond to the average speeds and average accelerations of the CMEs projected onto the sky plane as they pass within the LASCO field of view \citep{gopalswamy_2009}, which we note have been used to quantify CME speeds and accelerations in past analyses \citep[e.g.,][]{gopalswamy_2001}. Additionally, the CME masses were estimated using the procedure outlined in \citet{vourlidas_2000}. CDAW advises that their mass estimates only be taken as representative, and asterisks (*) adjacent to the CME masses we provide indicate that CDAW has identified some uncertainty in their estimation.\footnote{\url{https://cdaw.gsfc.nasa.gov/CME_list/catalog_description.htm}} For CMEs with no available mass estimate, the field is set to ``N/A''.

Of the 178 SAD events in the SADCat, there are seven pairs of SAD events associated with double-peak SXR flares where both flares correlate to the same CME and originate from the same source region. As such, there are 171 unique SAD-associated CMEs, 165 of which have available mass estimates. Because each entry in the catalog is based on an individual (single-peak) {GOES} SXR flare as recorded in the Hinode Flare Catalog, these pairs of events manifest as two separate entries in the SADCat with identical characteristic source region and CME values. To indicate that successive events share a CME, we include a dagger (\textdagger) next to both of their CME masses.

There is one event in the SADCat during which SADs are visible simultaneously above two different source regions. During event number 77 in Table~\ref{tbl:sads2}, SADs can be observed above sympathetic flaring in active region 11745 at N14~W87 and active region 11746 at S28~W76. The former's coordinates were used for flare--CME correlation, so they are the values provided in the SADCat.

\section{Discussion} \label{sec:discussion}

\begin{figure}
\centerline{\hspace*{0.015\textwidth}\includegraphics[scale=\figscale]{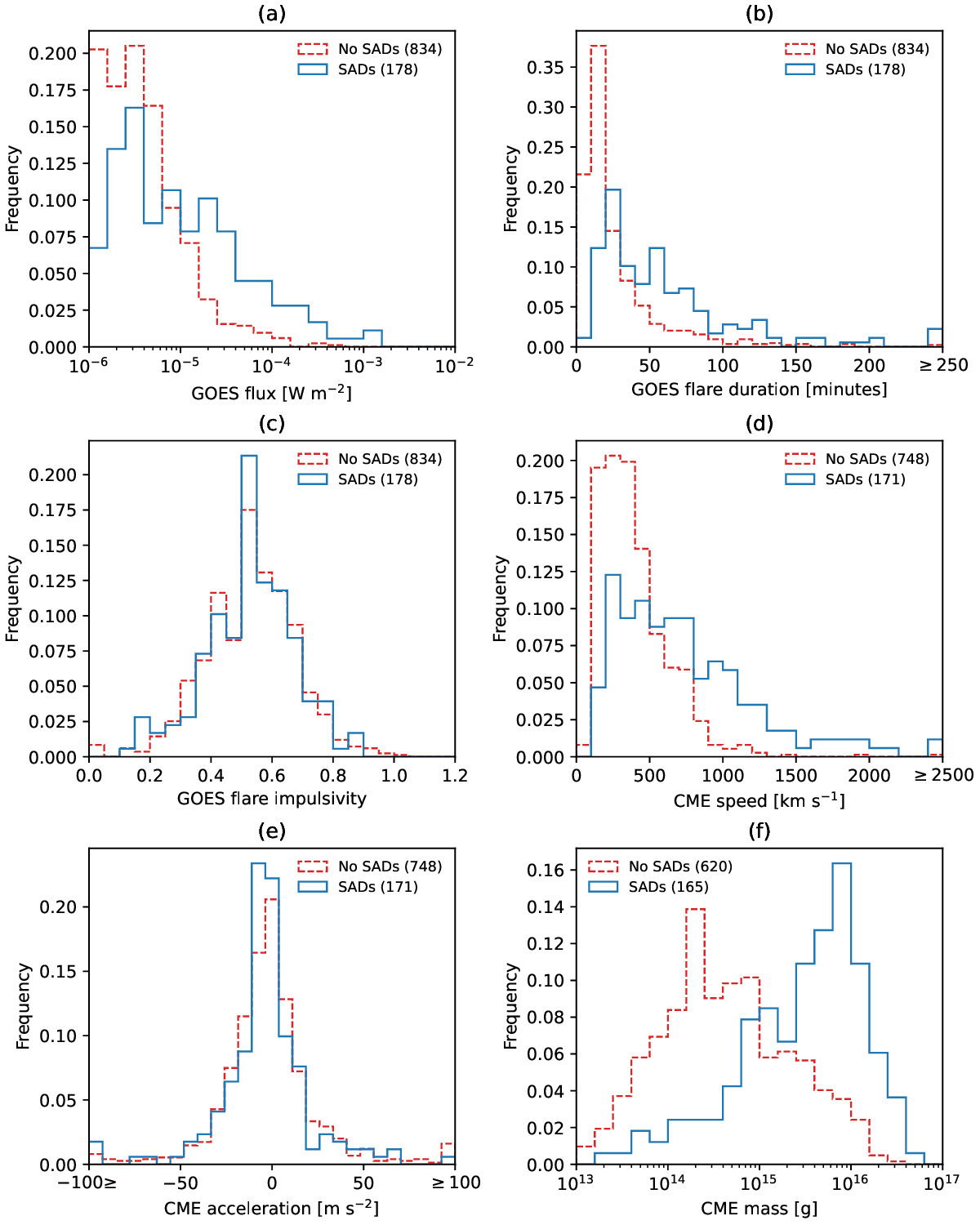}}
\caption{Frequency distributions of the (a) peak {GOES} flux, (b) flare duration, (c) flare impulsivity, (d) CME speed, (e) CME acceleration, and (f) CME mass for events without SADs (red dashed lines) and with SADs (blue solid lines).}
\label{fig:hist_freq}
\end{figure}

\begin{figure}
\centerline{\hspace*{0.015\textwidth}\includegraphics[scale=\figscale]{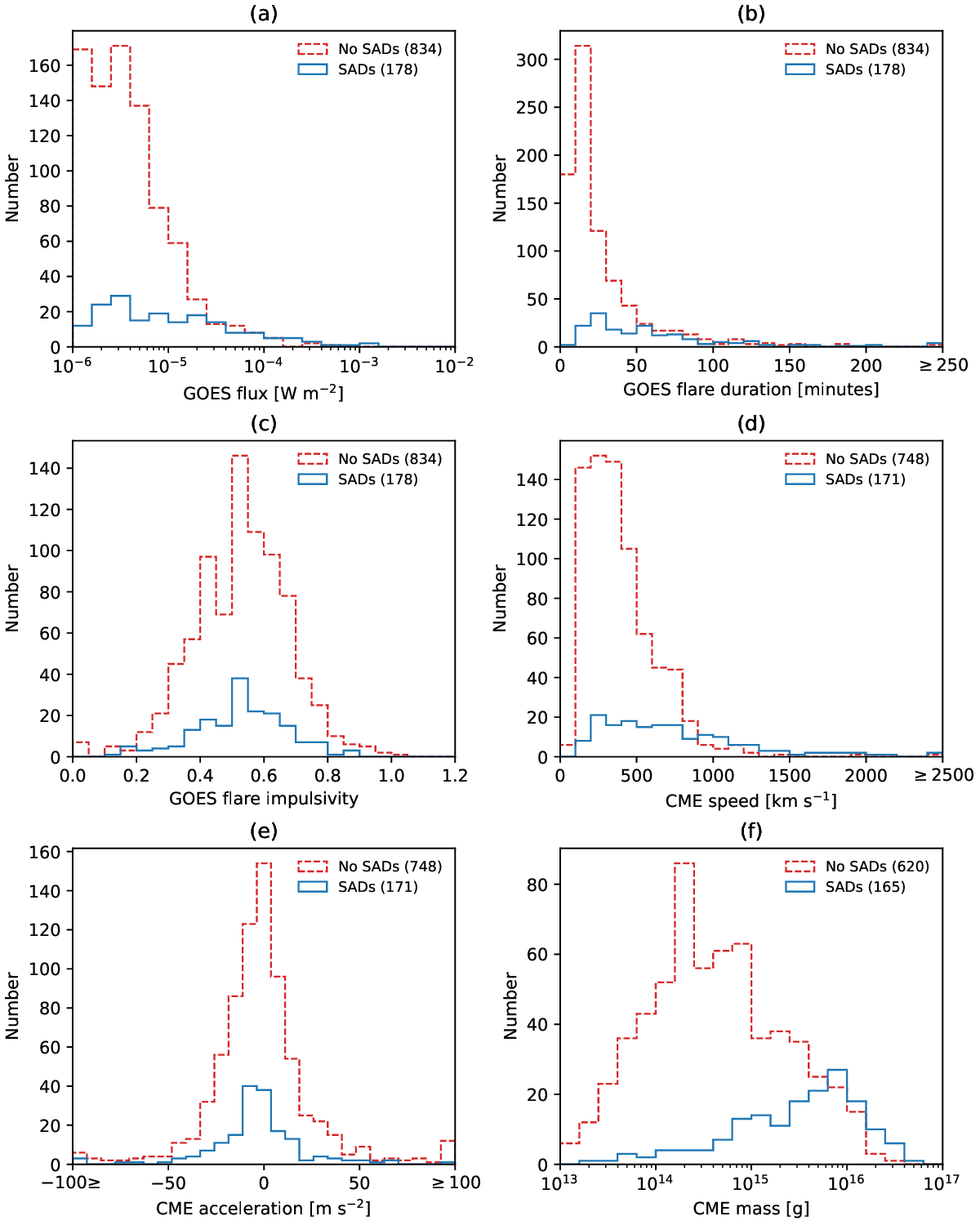}}
\caption{Number distributions of the (a) unsigned HS source-region longitude, (b) peak {GOES} flux, (c) flare duration, (d) flare impulsivity, (e) CME speed, and (f) CME mass for events without SADs (red dashed lines) and with SADs (blue solid lines).}
\label{fig:hist_num}
\end{figure}

For events with and without SADs, the distributions of the peak {GOES} flux, flare duration, flare impulsivity, CME speed, CME acceleration, and CME mass are provided in panels (a) to (f) of Figures~\ref{fig:hist_freq} and~\ref{fig:hist_num}. Figure~\ref{fig:hist_freq} plots the frequency distributions, while Figure~\ref{fig:hist_num} plots the number distributions. The frequency distributions can be obtained from the number distributions by dividing by the corresponding total number of flares or CMEs in the distribution as provided in each panel's legend. In panels (d) and (e), the distribution labeled ``No SADs'' consists of the 748 unique CMEs that correlate with exactly zero SAD-associated flares, and the distribution labeled ``SADs'' consists of the 171 unique CMEs that correlate with at least one SAD-associated flare. The distributions in panel (f) consist of the same CMEs as in (d) and (e) but excluding those with missing mass estimates giving a total of 620 CMEs not associated with any SADs and 165 CMEs associated with SADs.

From panels~(a) and~(b) in Figures~\ref{fig:hist_freq} and~\ref{fig:hist_num}, it is visually apparent that larger and longer-duration flares were more likely to produce SADs while the distributions of impulsivities in panel~(c) do not differ significantly. Peak flux was greater than $10^{-5}~{\rm W\ m^{-2}}$ (i.e., M or ``Moderate'' class and above) during 44\% (79/178) of the flares with SADs but only 15\% (127/834) of the flares without SADs. This is likely not a consequence of larger and longer flares producing coronal downflows, but rather a consequence of these events producing brighter supra-arcade fans where SADs are more easily detected. The average duration for SAD-associated flares is $62\pm4$~min (where error is the standard deviation of the mean), and the average for candidate flares not associated with SADs is $27\pm1$~min. Additionally, duration was greater than $30$~minutes, a rough cutoff for a so-called ``long-duration event'', for 65\% (116/178) of the flares with SADs but only 25\% (209/834) of the flares without SADs. To our knowledge, this is the first time this relationship between SAD occurrence and flare magnitude and duration has been demonstrated quantitatively. The average impulsivities are $0.526\pm 0.011$ and $0.523\pm 0.005$ for flares with and without SADs, respectively.

From panels (d) and (f), it is evident that SADs are more common for flares with faster and more massive CMEs despite similar acceleration distributions in panel (e). The average speed for SAD-associated CMEs is $745\pm 37\ {\rm km\ s^{-1}}$, and the average for those not associated with any SADs is $396\pm 9\ {\rm km\ s^{-1}}$. The speed was greater than 500~km~s$^{-1}$ for 63\% (108/171) of the CMEs with SADs and 25\% (189/748) of the CMEs without any SADs. The average accelerations for SAD-associated CMEs is $-5.1\pm 2.6\ {\rm m\ s}^{-2}$, and the average for candidate CMEs not associated with any SADs is $-1.1\pm 1.1\ {\rm m\ s}^{-2}$. The CME mass was greater than $10^{15}$~g in 76\% (126/165) of CMEs with SADs and 28\% (175/620) of the CMEs without any SADs.

The relationship between SAD occurrence and CME mass and speed is an interesting one, and not one that has been shown before. Offering a robust interpretation for this association is beyond the scope of this work, and likely requires MHD modeling. We do, however, note that the work of \citet{savage_2011} find individual SADs to be larger in the presence of weaker coronal magnetic fields. One theoretical interpretation could be that faster CMEs evacuate the corona faster, leaving a weaker magnetic field in their wake around the region surrounding the flare current sheet. This would perhaps contribute to a magnetic environment where SADs can become large enough to be detectable by current instrumentation.

\section{Summary and Conclusions} \label{sec:summary}
We have compiled a catalog (SADCat) of 178 SAD events observed by SDO/AIA during solar cycle~24, which we have made available in Table~\ref{tbl:sads1} and online at \sadcatlink. We intend for this catalog to be used by the solar physics community for further analysis of SAD events.

We have also examined the distributions of different values characterizing the associated flares and CMEs for the events in the SADCat. Some of the key findings are as follows.
\begin{itemize}
    \item Stronger and longer-duration solar flares are more likely to produce visible SADs, whereas the impulsivity of the flare has little impact.
    \item The occurrence of SADs is typically related to faster and more massive CMEs, but we find no significant association with CME acceleration.
\end{itemize}

To our knowledge, these relationships between SAD production in flares and flare/CME properties have not been reported quantitatively before. We note that future simulation and theoretical studies on the formation of SADs in flares should be compatible with the initial statistical results presented in this work. In the future, we plan to extend SADCat to cover events beyond solar cycle 24 and our selection criteria. With the dataset of SAD events collected in the current iteration of SADCat, it may be possible to apply machine-learning techniques to find additional events within our longitude and flux ranges. For events outside of these ranges, a citizen-science-based approach may be fruitful.


\newcommand{\tablefontsize}{\tiny}
\newcommand{\tableheader}{\hline
\multicolumn{1}{r}{Event} & \multicolumn{4}{l}{{GOES} SXR flare} & \multicolumn{3}{l}{Source region} & \multicolumn{2}{l}{CME} \\
No. & Peak time & Dur. & Imp. & Class & No. & Lat. & Lon. & Speed & Mass \\
& [UTC] & [min] & & & & [\degree] & [\degree] & [$\rm km/s$] & [g]
\\
\hline}
\begin{table}[h]
\tablefontsize
\begin{tabular}{rlrrlrrrrr@{}l}
\caption{
	SADCat events numbered 1 to 57.
Described fully in Section~\ref{sec:results}.
}
\label{tbl:sads1} \\
\tableheader
1 & 2010-06-13 05:39 & 14 & 0.64 & M1.5 & 11079 & S24 & W82 & 320 & 8.7E+14 &  \\
2 & 2010-08-18 05:48 & 126 & 0.50 & C6.5 & 11099 & N18 & W88 & 1471 & 1.1E+16 & * \\
3 & 2010-09-08 23:28 & 28 & 0.82 & C4.7 & 11105 & N21 & W87 & 818 & 6.3E+15 & * \\
4 & 2010-11-03 06:11 & 32 & 0.53 & C5.5 & 11121 & S19 & E88 & 270 & 7.6E+14 &  \\
5 & 2011-01-14 03:30 & 190 & 0.64 & C2.4 & 11147 & N23 & E87 & 625 & 1.6E+15 &  \\
6 & 2011-01-28 01:03 & 121 & 0.16 & M2.0 & 11149 & N16 & W88 & 606 & 3.5E+15 &  \\
7 & 2011-02-11 21:57 & 56 & 0.46 & C1.1 & 11160 & N12 & E88 & 469 & 2.6E+15 &  \\
8 & 2011-02-24 07:35 & 19 & 0.63 & M5.0 & 11163 & N15 & E87 & 1186 & 7.5E+15 & * \\
9 & 2011-03-08 03:45 & 21 & 0.38 & M2.1 & 11171 & S21 & E72 & 732 & 6.0E+15 & * \\
10 & 2011-03-08 18:28 & 33 & 0.61 & M6.4 & 11165 & S17 & W88 & 283 & 3.7E+15 &  \\
11 & 2011-03-08 20:16 & 120 & 0.34 & M2.1 & 11165 & S19 & W87 & 702 & 5.4E+15 &  \\
12 & 2011-03-09 02:31 & 6 & 0.17 & C9.2 & 11171 & S26 & W70 & 234 & 1.2E+14 &  \\
13 & 2011-04-18 04:03 & 87 & 0.52 & C2.6 & 11186 & N26 & W87 & 250 & 9.3E+14 &  \\
14 & 2011-05-09 20:59 & 37 & 0.46 & C7.8 & N/A & N16 & E88 & 1318 & 1.0E+16 & * \\
15 & 2011-05-18 18:30 & 58 & 0.48 & C2.9 & 11208 & N13 & W89 & 1105 & 6.3E+15 & * \\
16 & 2011-06-29 00:32 & 37 & 0.32 & C1.0 & 11240 & S19 & W81 & 481 & 2.7E+15 & * \\
17 & 2011-08-06 18:07 & 106 & 0.68 & C2.0 & 11260 & N25 & W87 & 176 & 5.1E+14 &  \\
18 & 2011-08-09 03:45 & 35 & 0.74 & M3.6 & 11263 & N17 & W69 & 1146 & 1.2E+15 & * \\
19 & 2011-08-09 08:05 & 20 & 0.85 & X10 & 11263 & N14 & W69 & 1610 & 1.1E+16 & * \\
20 & 2011-08-11 10:23 & 62 & 0.79 & C8.9 & 11263 & N15 & W88 & 1160 & 3.3E+15 & * \\
21 & 2011-08-17 01:54 & 70 & 0.14 & C2.8 & 11272 & S19 & E72 & 550 & 3.3E+15 &  \\
22 & 2011-09-12 20:54 & 81 & 0.30 & M1.4 & 11295 & N23 & E87 & 154 & 4.4E+13 &  \\
23 & 2011-09-13 13:02 & 74 & 0.80 & C3.8 & 11283 & N20 & W87 & 746 & 6.9E+15 & * \\
24 & 2011-09-22 11:01 & 75 & 0.43 & X2.1 & 11302 & N09 & E89 & 1905 & 2.1E+16 & * \\
25 & 2011-10-22 11:10 & 189 & 0.37 & M1.9 & 11314 & N27 & W87 & 1005 & 1.2E+16 & * \\
26 & 2011-11-04 01:01 & 77 & 0.45 & C7.7 & 11339 & N22 & E88 & 756 & 1.1E+16 & * \\
27 & 2011-11-13 08:20 & 23 & 0.39 & C2.7 & 11344 & S20 & W72 & 158 & 5.0E+13 &  \\
28 & 2011-11-14 20:13 & 52 & 0.60 & C2.5 & 11348 & N18 & W70 & 383 & 8.4E+14 &  \\
29 & 2011-11-14 23:58 & 25 & 0.64 & C5.8 & 11339 & N21 & W72 & 285 & 9.3E+14 &  \\
30 & 2011-11-17 20:11 & 30 & 0.23 & C3.4 & 11353 & N13 & E89 & 1041 & 1.4E+16 & * \\
31 & 2011-11-18 02:30 & 58 & 0.45 & C3.8 & 11348 & N24 & W89 & 601 & 3.4E+15 &  \\
32 & 2011-11-30 20:28 & 19 & 0.42 & C5.4 & 11364 & N21 & E67 & 716 & 6.4E+15 & * \\
33 & 2011-11-30 22:32 & 55 & 0.71 & C5.2 & 11356 & N13 & W67 & 634 & 7.2E+15 & * \\
34 & 2011-12-03 11:22 & 110 & 0.38 & C2.3 & 11355 & N21 & W89 & 645 & 5.4E+15 & * \\
35 & 2011-12-11 08:27 & 60 & 0.45 & C1.3 & 11375 & N21 & E84 & 398 & 1.1E+15 &  \\
36 & 2011-12-24 08:31 & 21 & 0.19 & C7.5 & 11386 & S18 & E70 & 732 & 1.5E+15 & * \\
37 & 2012-01-16 04:44 & 250 & 0.51 & C9.3 & 11402 & N34 & E86 & 1060 & 1.0E+16 & * \\
38 & 2012-01-20 21:12 & 21 & 0.52 & C2.0 & 11396 & N25 & W71 & 354 & 1.1E+15 &  \\
39 & 2012-01-21 13:42 & 12 & 0.58 & C3.4 & 11396 & N25 & W82 & 377 & 7.5E+14 &  \\
40 & 2012-01-27 18:37 & 79 & 0.76 & X2.6 & 11402 & N33 & W85 & 2508 & 3.7E+16 & * \\
41 & 2012-03-04 10:52 & 107 & 0.21 & M2.9 & 11429 & N16 & E65 & 1306 & 7.9E+15 & * \\
42 & 2012-03-09 20:25 & 50 & 0.52 & M1.4 & 11432 & N17 & E79 & 598 & 6.2E+15 &  \\
43 & 2012-03-13 17:30 & 29 & 0.62 & X1.1 & 11430 & N17 & W66 & 1884 & 2.3E+16 & * \\
44 & 2012-03-17 23:28 & 21 & 0.38 & C1.1 & 11429 & N19 & W87 & 1210 & 1.1E+16 & * \\
45 & 2012-04-09 12:44 & 56 & 0.57 & C5.7 & 11451 & N20 & W65 & 921 & 7.5E+15 & * \\
46 & 2012-04-15 01:40 & 68 & 0.63 & C3.9 & 11458 & N11 & E88 & 1220 & 5.8E+15 & *\textdagger \\
47 & 2012-04-15 02:34 & 29 & 0.62 & C2.5 & 11458 & N15 & E88 & 1220 & 5.8E+15 & *\textdagger \\
48 & 2012-04-16 00:46 & 44 & 0.50 & C2.7 & 11461 & N12 & E88 & 1128 & 1.4E+15 &  \\
49 & 2012-04-16 17:40 & 36 & 0.44 & M2.5 & 11461 & N14 & E88 & 1348 & 7.3E+15 & * \\
50 & 2012-04-17 08:09 & 166 & 0.74 & C1.6 & 11461 & N13 & E88 & 533 & N/A &  \\
51 & 2012-05-03 14:32 & 107 & 0.64 & C2.3 & 11475 & N09 & E89 & 584 & 6.3E+15 & * \\
52 & 2012-05-17 01:47 & 49 & 0.45 & M7.3 & 11476 & N07 & W88 & 1582 & 3.6E+16 & * \\
53 & 2012-06-01 22:41 & 32 & 0.34 & C4.8 & 11488 & N12 & W89 & 630 & 4.2E+15 & * \\
54 & 2012-06-23 07:50 & 102 & 0.47 & C3.9 & 11506 & N14 & W89 & 1263 & 1.0E+16 & * \\
55 & 2012-07-08 16:32 & 19 & 0.47 & M9.9 & 11515 & S17 & W74 & 822 & 5.9E+15 & * \\
56 & 2012-07-17 17:15 & 421 & 0.74 & M2.5 & 11520 & S28 & W65 & 958 & 1.7E+16 & * \\
57 & 2012-07-19 05:58 & 159 & 0.64 & X1.1 & 11520 & S13 & W88 & 1631 & 2.1E+16 & * \\
\hline
\end{tabular}
\end{table}
\setcounter{table}{0}
\begin{table}[h]
\tablefontsize
\begin{tabular}{rlrrlrrrrr@{}l}
\caption{
	\textbf{(continued)} SADCat events numbered 58 to 114.
}
\label{tbl:sads2} \\
\tableheader
58 & 2012-08-02 13:10 & 85 & 0.71 & C2.3 & 11529 & S20 & W87 & 563 & 7.2E+15 &  \\
59 & 2012-08-13 09:30 & 22 & 0.50 & C2.1 & 11538 & S25 & W86 & 600 & 1.8E+14 &  \\
60 & 2012-08-18 01:02 & 43 & 0.88 & M8.0 & 11548 & N19 & E86 & 986 & 3.6E+15 & * \\
61 & 2012-08-27 10:18 & 44 & 0.55 & C1.1 & 11556 & S14 & W77 & 194 & 9.0E+14 &  \\
62 & 2012-10-07 20:46 & 53 & 0.47 & C1.8 & 11589 & N21 & E87 & 692 & 2.0E+15 & * \\
63 & 2012-10-18 22:59 & 51 & 0.57 & C2.0 & 11596 & N11 & E89 & 460 & 3.3E+14 &  \\
64 & 2012-10-28 08:05 & 262 & 0.38 & C2.4 & N/A & S26 & W87 & 143 & 1.1E+14 &  \\
65 & 2012-11-11 02:33 & 41 & 0.54 & M1.5 & 11614 & N15 & E89 & 407 & 5.0E+14 &  \\
66 & 2012-12-05 00:28 & 38 & 0.42 & C2.5 & 11628 & N14 & E83 & 963 & 3.2E+15 & * \\
67 & 2012-12-25 10:39 & 26 & 0.50 & C1.7 & 11639 & S13 & E89 & 197 & 2.7E+13 &  \\
68 & 2013-01-05 00:18 & 93 & 0.34 & C3.4 & 11652 & N21 & E88 & 269 & 2.1E+13 &  \\
69 & 2013-02-14 04:30 & 26 & 0.69 & C1.5 & 11667 & N21 & W87 & 851 & 1.4E+15 &  \\
70 & 2013-04-05 06:50 & 125 & 0.35 & C3.0 & 11719 & N11 & E88 & 588 & 9.5E+15 & * \\
71 & 2013-04-18 18:23 & 68 & 0.40 & C9.3 & 11719 & N11 & W88 & 495 & 5.3E+15 & * \\
72 & 2013-05-03 17:32 & 17 & 0.47 & M8.2 & 11739 & N16 & E81 & 858 & 9.0E+15 & * \\
73 & 2013-05-11 19:48 & 28 & 0.46 & M1.1 & 11746 & S28 & E73 & 116 & 6.7E+13 &  \\
74 & 2013-05-13 02:17 & 39 & 0.62 & X2.5 & 11748 & N11 & E89 & 1270 & 1.6E+16 & * \\
75 & 2013-05-13 16:01 & 49 & 0.27 & X4.1 & 11748 & N11 & E85 & 1850 & 1.1E+16 & * \\
76 & 2013-05-20 05:25 & 47 & 0.19 & M2.5 & 11755 & N09 & E89 & 334 & 7.7E+15 & * \\
77 & 2013-05-22 13:32 & 60 & 0.40 & M7.2 & 11745 & N14 & W87 & 1466 & 3.3E+16 & * \\
78 & 2013-05-31 08:07 & 104 & 0.63 & C1.9 & 11756 & S21 & W89 & 660 & 3.6E+15 &  \\
79 & 2013-06-14 00:31 & 82 & 0.46 & C1.7 & 11775 & S21 & E72 & 273 & 2.7E+15 &  \\
80 & 2013-06-18 15:57 & 60 & 0.40 & C1.6 & 11768 & S11 & W89 & 279 & 7.6E+14 & \textdagger \\
81 & 2013-06-18 16:54 & 24 & 0.79 & C1.3 & 11768 & S13 & W89 & 279 & 7.6E+14 & \textdagger \\
82 & 2013-06-21 03:14 & 73 & 0.60 & M4.2 & 11777 & S16 & E73 & 1900 & 8.7E+15 & * \\
83 & 2013-07-02 23:58 & 22 & 0.59 & C9.5 & 11785 & S14 & E88 & 592 & 2.5E+15 & * \\
84 & 2013-07-03 07:08 & 18 & 0.44 & M2.2 & 11787 & S11 & E82 & 807 & 3.4E+15 & * \\
85 & 2013-07-04 11:41 & 16 & 0.44 & C2.6 & 11787 & S12 & E66 & 188 & 5.7E+14 & \textdagger \\
86 & 2013-07-04 12:17 & 10 & 0.50 & C2.4 & 11787 & S13 & E68 & 188 & 5.7E+14 & \textdagger \\
87 & 2013-07-18 18:23 & 62 & 0.44 & C3.3 & 11800 & S12 & E81 & 458 & N/A &  \\
88 & 2013-09-06 22:50 & 57 & 0.60 & C1.4 & N/A & N13 & E88 & 348 & N/A &  \\
89 & 2013-10-02 21:51 & 206 & 0.57 & C2.1 & 11850 & N19 & W87 & 619 & 9.7E+15 & * \\
90 & 2013-10-11 07:25 & 44 & 0.55 & M2.2 & 11869 & N21 & E87 & 1200 & 1.0E+16 & * \\
91 & 2013-10-14 03:24 & 11 & 0.64 & C1.6 & 11869 & N15 & E83 & 210 & 8.4E+13 &  \\
92 & 2013-10-14 21:27 & 14 & 0.50 & C4.8 & 11861 & N10 & E89 & 356 & 2.2E+15 & *\textdagger \\
93 & 2013-10-14 21:53 & 25 & 0.60 & C4.3 & 11861 & N10 & E89 & 356 & 2.2E+15 & *\textdagger \\
94 & 2013-10-25 03:02 & 24 & 0.58 & M4.3 & 11882 & S07 & E76 & 344 & 7.0E+15 & * \\
95 & 2013-10-25 08:01 & 16 & 0.50 & X2.5 & 11882 & S08 & E73 & 587 & 5.2E+15 & * \\
96 & 2013-10-25 15:03 & 21 & 0.57 & X3.1 & 11882 & S06 & E69 & 1081 & 7.1E+15 & * \\
97 & 2013-11-05 14:22 & 52 & 0.50 & C3.6 & 11888 & S13 & W88 & 468 & 1.3E+15 &  \\
98 & 2013-11-07 00:02 & 30 & 0.60 & M2.7 & 11882 & S11 & W88 & 1033 & 7.3E+15 & * \\
99 & 2013-11-21 11:11 & 50 & 0.38 & M1.8 & 11893 & S14 & W89 & 668 & 3.8E+15 & * \\
100 & 2013-11-25 11:49 & 134 & 0.80 & C2.9 & 11904 & N15 & W89 & 571 & 2.1E+14 &  \\
101 & 2013-12-16 21:33 & 76 & 0.53 & C4.4 & 11927 & S26 & W89 & 779 & 8.8E+15 & * \\
102 & 2013-12-24 12:59 & 39 & 0.44 & C1.7 & 11928 & S15 & W89 & 730 & 9.6E+14 &  \\
103 & 2013-12-31 11:50 & 129 & 0.67 & M1.3 & 11943 & S09 & E89 & 1101 & 4.1E+15 & * \\
104 & 2014-01-04 22:52 & 70 & 0.57 & M2.8 & 11936 & S14 & W89 & 567 & 3.6E+15 & * \\
105 & 2014-01-08 03:47 & 15 & 0.53 & M5.3 & 11947 & N11 & W81 & 643 & 2.5E+15 &  \\
106 & 2014-01-16 21:39 & 32 & 0.56 & C9.0 & 11959 & S23 & E88 & 438 & 2.1E+15 &  \\
107 & 2014-01-17 20:07 & 74 & 0.68 & M1.0 & 11959 & S23 & E88 & 721 & 3.2E+15 & * \\
108 & 2014-01-26 08:38 & 67 & 0.18 & C2.2 & 11967 & S17 & E88 & 1088 & 1.7E+15 & * \\
109 & 2014-01-27 01:22 & 34 & 0.50 & M1.5 & 11967 & S16 & E88 & 300 & 5.3E+14 &  \\
110 & 2014-01-27 02:11 & 16 & 0.56 & M1.6 & 11967 & S13 & E88 & 687 & 8.7E+14 & * \\
111 & 2014-02-09 16:17 & 72 & 0.51 & M1.5 & 11977 & S16 & E88 & 908 & 7.2E+15 & * \\
112 & 2014-02-20 07:56 & 59 & 0.51 & M4.4 & 11976 & S15 & W73 & 948 & 8.7E+15 & * \\
113 & 2014-02-24 11:17 & 39 & 0.36 & M1.7 & 11990 & S11 & E88 & 495 & 7.0E+15 & * \\
114 & 2014-02-25 00:49 & 24 & 0.42 & X7.1 & 11990 & S12 & E82 & 2147 & 2.2E+16 & * \\
\hline
\end{tabular}
\end{table}
\setcounter{table}{0}
\begin{table}[h]
\tablefontsize
\begin{tabular}{rlrrlrrrrr@{}l}
\caption{
	\textbf{(continued)} SADCat events numbered 115 to 171.
}
\label{tbl:sads3} \\
\tableheader
115 & 2014-04-11 15:01 & 66 & 0.71 & C7.6 & 12035 & S15 & E88 & 742 & 5.7E+15 & * \\
116 & 2014-04-12 07:27 & 52 & 0.23 & C7.2 & 12035 & S14 & E88 & 1016 & 5.2E+14 & * \\
117 & 2014-04-23 23:53 & 43 & 0.72 & C4.7 & 12036 & S21 & W88 & 680 & 1.6E+15 &  \\
118 & 2014-04-25 00:27 & 21 & 0.48 & X2.0 & 12035 & S14 & W89 & 456 & 4.6E+15 & * \\
119 & 2014-05-06 17:35 & 70 & 0.36 & C6.7 & 12051 & S09 & W89 & 815 & 7.0E+15 & * \\
120 & 2014-05-07 16:29 & 56 & 0.39 & M1.7 & 12051 & S11 & W89 & 923 & 1.3E+16 & * \\
121 & 2014-05-18 06:54 & 168 & 0.26 & C5.4 & 12056 & N06 & W89 & 324 & 7.8E+14 &  \\
122 & 2014-06-10 12:52 & 27 & 0.59 & X2.2 & 12087 & S17 & E82 & 1469 & 1.7E+16 & * \\
123 & 2014-06-27 01:14 & 337 & 0.62 & C2.3 & 12100 & N13 & E89 & 497 & 9.1E+15 & * \\
124 & 2014-07-03 03:59 & 52 & 0.52 & C3.7 & N/A & S22 & E88 & 1054 & 6.6E+14 &  \\
125 & 2014-08-28 15:20 & 47 & 0.26 & C2.7 & 12146 & N08 & W77 & 433 & 1.3E+15 &  \\
126 & 2014-09-03 13:54 & 63 & 0.54 & M3.6 & 12157 & S15 & E87 & 468 & 1.1E+15 &  \\
127 & 2014-09-04 16:32 & 27 & 0.41 & C8.2 & 12155 & S19 & E87 & 240 & 2.9E+14 &  \\
128 & 2014-09-05 06:53 & 53 & 0.70 & C9.6 & 12157 & S14 & E69 & 740 & 3.7E+15 & * \\
129 & 2014-10-02 17:44 & 65 & 0.52 & M2.3 & 12172 & S18 & W76 & 217 & 3.2E+15 &  \\
130 & 2014-10-02 19:01 & 25 & 0.48 & X1.0 & 12173 & S17 & W82 & 513 & 1.1E+16 & * \\
131 & 2014-10-14 18:37 & 25 & 0.64 & M1.6 & 12192 & S12 & E88 & 848 & 3.0E+16 & * \\
132 & 2014-10-22 15:57 & 12 & 0.50 & M2.0 & 12192 & S11 & E88 & 434 & 1.9E+15 &  \\
133 & 2014-10-22 21:16 & 21 & 0.67 & C9.4 & 12192 & S14 & E88 & 317 & 7.2E+14 &  \\
134 & 2014-11-02 21:05 & 41 & 0.51 & M1.4 & 12205 & N15 & E89 & 289 & 1.2E+15 & * \\
135 & 2014-11-03 11:53 & 54 & 0.56 & M3.2 & 12205 & N17 & E89 & 447 & 8.6E+15 & * \\
136 & 2014-11-04 08:38 & 52 & 0.75 & M3.8 & 12205 & N15 & E82 & 627 & 5.0E+15 & *\textdagger \\
137 & 2014-11-04 09:18 & 53 & 0.68 & M3.5 & 12205 & N15 & E82 & 627 & 5.0E+15 & *\textdagger \\
138 & 2014-11-05 19:44 & 85 & 0.64 & M4.3 & 12205 & N17 & E65 & 608 & 1.1E+16 & * \\
139 & 2014-11-06 14:24 & 16 & 0.44 & C5.5 & 12208 & S12 & E88 & 798 & 1.1E+15 &  \\
140 & 2014-12-05 06:07 & 20 & 0.50 & C2.6 & 12226 & S21 & W69 & 534 & 6.3E+15 & * \\
141 & 2014-12-10 19:22 & 203 & 0.67 & C8.4 & 12222 & S18 & W89 & 1086 & 7.8E+15 & * \\
142 & 2014-12-12 23:13 & 19 & 0.53 & C3.7 & 12241 & S09 & E89 & 278 & 1.8E+14 &  \\
143 & 2015-01-13 04:24 & 25 & 0.44 & M8.1 & 12257 & N06 & W70 & 525 & N/A & \textdagger \\
144 & 2015-01-13 04:58 & 24 & 0.50 & M7.1 & 12257 & N05 & W76 & 525 & N/A & \textdagger \\
145 & 2015-02-24 10:57 & 119 & 0.53 & C2.1 & N/A & S21 & E87 & 902 & 1.1E+16 & * \\
146 & 2015-03-01 16:13 & 58 & 0.69 & C9.8 & 12290 & N20 & W74 & 387 & 1.1E+14 &  \\
147 & 2015-03-02 15:28 & 27 & 0.67 & M5.4 & 12290 & N21 & W86 & 452 & 1.6E+15 & * \\
148 & 2015-03-02 19:31 & 15 & 0.67 & M5.9 & 12290 & N19 & W87 & 227 & 5.0E+13 &  \\
149 & 2015-03-03 00:27 & 17 & 0.59 & C5.1 & 12290 & N19 & W87 & 344 & 1.3E+15 & \textdagger \\
150 & 2015-03-03 01:35 & 17 & 0.59 & X1.2 & 12290 & N21 & W87 & 344 & 1.3E+15 & \textdagger \\
151 & 2015-03-06 08:15 & 93 & 0.86 & M2.2 & 12297 & S20 & E87 & 880 & 8.4E+15 & * \\
152 & 2015-03-07 22:22 & 73 & 0.51 & X1.3 & 12297 & S19 & E74 & 1261 & 2.7E+16 & * \\
153 & 2015-04-03 07:26 & 67 & 0.52 & C2.4 & 12318 & S01 & E81 & 429 & 4.1E+15 & * \\
154 & 2015-04-11 21:00 & 96 & 0.48 & C2.6 & 12321 & N10 & E80 & 224 & 1.4E+14 &  \\
155 & 2015-04-21 10:40 & 42 & 0.55 & M3.2 & 12333 & N16 & E80 & 2039 & 4.5E+15 & * \\
156 & 2015-04-23 10:07 & 112 & 0.44 & M1.6 & 12321 & N07 & W80 & 857 & 2.1E+16 & * \\
157 & 2015-05-05 22:11 & 10 & 0.60 & X3.9 & 12339 & N15 & E79 & 715 & 3.9E+16 & * \\
158 & 2015-05-12 03:02 & 87 & 0.54 & C3.7 & 12335 & S21 & W83 & 772 & 1.9E+16 & * \\
159 & 2015-06-04 09:47 & 20 & 0.55 & M1.2 & 12361 & N18 & E73 & 206 & 2.6E+14 &  \\
160 & 2015-06-18 01:27 & 82 & 0.66 & M1.8 & 12365 & S16 & W81 & 1714 & 1.0E+16 & * \\
161 & 2015-07-18 14:42 & 24 & 0.58 & C2.6 & 12388 & N14 & W80 & 329 & 1.6E+15 & * \\
162 & 2015-08-02 22:43 & 134 & 0.62 & C1.1 & N/A & S24 & E89 & 460 & 1.1E+16 & * \\
163 & 2015-09-29 05:37 & 6 & 0.67 & M1.7 & 12423 & S09 & W82 & 503 & 2.2E+14 &  \\
164 & 2015-10-01 05:53 & 37 & 0.38 & C2.3 & 12428 & S04 & W89 & 914 & 3.3E+15 &  \\
165 & 2015-11-07 05:32 & 19 & 0.47 & C3.4 & 12450 & S25 & E78 & 243 & 6.1E+14 &  \\
166 & 2015-12-22 03:34 & 33 & 0.58 & M2.4 & 12473 & S23 & E75 & 258 & 2.8E+14 &  \\
167 & 2016-01-02 00:11 & 111 & 0.55 & M3.4 & 12473 & S25 & W82 & 1730 & 7.3E+15 & * \\
168 & 2016-01-29 21:46 & 129 & 0.45 & C2.8 & 12487 & S20 & W83 & 901 & 8.4E+15 &  \\
169 & 2016-03-10 18:54 & 24 & 0.50 & C1.2 & 12512 & N14 & W87 & 721 & 1.2E+15 &  \\
170 & 2016-03-16 06:46 & 23 & 0.52 & C3.2 & 12522 & N12 & W88 & 592 & 4.2E+15 & * \\
171 & 2016-07-24 17:43 & 42 & 0.31 & M2.8 & 12567 & N07 & W89 & 244 & N/A &  \\
\hline
\end{tabular}
\end{table}
\setcounter{table}{0}
\begin{table}[h]
\tablefontsize
\begin{tabular}{rlrrlrrrrr@{}l}
\caption{
	\textbf{(continued)} SADCat events numbered 172 to 178.
}
\label{tbl:sads4} \\
\tableheader
172 & 2016-12-10 17:15 & 49 & 0.55 & C5.7 & 12615 & S07 & W89 & 248 & 1.2E+15 &  \\
173 & 2017-04-18 20:10 & 88 & 0.56 & C7.9 & 12651 & N14 & E77 & 926 & 2.0E+16 & * \\
174 & 2017-06-01 07:53 & 76 & 0.79 & C2.0 & 12661 & N05 & E88 & 285 & 5.0E+14 &  \\
175 & 2017-09-02 15:41 & 30 & 0.50 & M1.2 & 12672 & N04 & W89 & 705 & 5.8E+15 &  \\
176 & 2017-09-09 23:53 & 157 & 0.69 & M1.6 & 12673 & S07 & W74 & 1019 & 1.1E+16 & * \\
177 & 2017-09-10 16:06 & 56 & 0.55 & X12 & 12673 & S08 & W88 & 3163 & 4.8E+16 & * \\
178 & 2017-10-20 23:28 & 27 & 0.67 & M1.6 & 12685 & S12 & E88 & 331 & N/A &  \\
\hline
\end{tabular}
\end{table}

\begin{acks}
This work was carried out using the Hinode Flare Catalogue (\url{https://hinode.isee.nagoya-u.ac.jp/flare_catalogue/}), which is maintained by ISAS/JAXA and Institute for Space-Earth Environmental Research (ISEE), Nagoya University. The SOHO/LASCO CME catalog used in this work is generated and maintained at the CDAW Data Center by NASA and The Catholic University of America in cooperation with the Naval Research Laboratory. SOHO is a project of international cooperation between ESA and NASA. The SDO data used is courtesy of NASA/SDO and the AIA, EVE, and HMI science teams.
\end{acks}

\begin{authorcontribution}
T.F.S. wrote the code, compiled the catalog, and created all figures and tables. R.J.F. conceived the study and provided valuable advice, feedback, and direction.
\end{authorcontribution}

\begin{fundinginformation}
\emergencystretch 3em
T.F.S received support from the Undergraduate Research Opportunities Program at the University of Colorado Boulder. R.J.F. received support from NASA HGI award 80NSSC25K7927.
\end{fundinginformation}

\begin{dataavailability}
Solar flare events used in this study were compiled from the Hinode Flare Catalogue at \url{https://hinode.isee.nagoya-u.ac.jp/flare_catalogue/}, CME events used were compiled from the {SOHO}/LASCO CME Catalog at \url{https://cdaw.gsfc.nasa.gov/CME_list/index.html}, and the {SDO}/AIA data were downloaded from the Joint Science Operations Center at \url{https://solarweb1.stanford.edu/}. The full SADCat is available at \sadcatlink.
\end{dataavailability}

\bibliographystyle{spr-mp-sola}
\bibliography{refs}

\end{document}